\documentclass[12pt]{article}

\usepackage{citesort}

\newcommand{\ice}[1]{\relax}

\newcommand{\re}[1]{(\ref{#1})}

\catcode`\@=11
\def\slash{\mathpalette\make@slash}
\def\make@slash#1#2{\setbox\z@\hbox{$#1#2$}%
  \hbox to 0pt{\hss$#1/$\hss\kern-\wd0}\box0}
\catcode`\@=12 % @ signs are no longer letters

\begin{document}

\begin{titlepage}
\noindent
%
% Datum
%
\hfill TTP-04--08\\
\mbox{}
\hfill SFB/CPP-04--16\\
\mbox{}
\hfill  May 2004  \\   
\mbox{}
\hfill hep-ph/0405193\\
\mbox{}
\hfill {}

\vspace{0.5cm}
\begin{center}
  \begin{Large}
  \begin{bf}
%Four-loop anomalous dimensions of gluon and ghost fields in QCD
Four-loop renormalization  of QCD: full set of renormalization  constants 
and  anomalous dimensions
  \end{bf}
  \end{Large}

%
% Author
%
  \vspace{0.8cm}

  \begin{large}
K.G. Chetyrkin\footnote{On leave from Institute for Nuclear Research
of the Russian Academy of Sciences, Moscow, 117312, Russia.} \\
{\small {\em
    Institut f\"ur Theoretische Teilchenphysik,
    Universit\"at Karlsruhe,
     D-76128 Karlsruhe, Germany}
}
 \\[3mm]

%[-5mm]
\end{large}

\vspace{0.8cm}
{\bf Abstract}
\end{center}
\begin{quotation}
\noindent
The anomalous dimensions of the gluon and ghost fields 
as well as those of the
ghost-ghost-gluon and quark-quark-gluon vertexes are analytically
computed at four loops in pQCD.  Taken together with already available anomalous
dimensions of the coupling constant, the quark field and the mass the
results lead to complete knowledge of {\em all} renormalization
constant entering into the renormalization of the QCD Lagrangian at
the four-loop level. As a by-product we get scale and scheme invariant
gluon and ghost propagators at NNNLO. Using a theorem due to Dudal,
Verschelde and Sorella, we also construct the four-loop anomalous
dimension of the ``gluon mass operator'', $A^2$, in the Landau gauge.
\end{quotation}
\end{titlepage}

\section{\label{intro} Introduction}

The  renormalization group equation is a powerful  tool in
investigating the properties of the Green functions of a
renormalizable field theory. Its crucial ingredients 
are the  anomalous dimensions of quantum fields   
as well as those of mass and coupling constant(s).

In recent years there  has been achieved a significant progress in
perturbative calculation of higher orders corrections to
renormalization group functions For example, the most physically
important RG functions of QCD---the  $\beta$ function and the quark mass and
field anomalous dimensions---have been computed at a record-setting
four-loup level 
\cite{vanRitbergen:1997va,Chetyrkin:1997dh,Vermaseren:1997fq,Chetyrkin:1999pq}.

In the same time the anomalous dimensions of the gluon and ghost
fields are available in literature only at three-loop level
\cite{Larin:1993tp,Chetyrkin:1999pq}. It is unfortunate for at  least two reasons. First, 
the gluon and    ghost  field anomalous dimensions are  important in
comparison of the non-perturbative 
results for the momentum
dependence of the corresponding  propagators with  perturbative predictions 
\cite{Suman:1996zg,%
Becirevic:1999uc,Becirevic:1999sc,Becirevic:1999hj,Nakajima:2000yf,Boucaud:2000nd,Boucaud:2001st,%
Boucaud:2001un,Bonnet:2001uh,Cucchieri:2001za,VanAcoleyen:2002nd,Furui:2003jr}.  
Second, only the knowledge of anomalous dimensions of {\em all} fields
of the QCD Lagrangian leads, together with the $\beta$- function, to
complete reconstruction of {\em all} Renormalization Constants (RCs)
entering into the renormalization of the QCD Lagrangian (see below).

In the present paper we fill the gap by analytically computing the
anomalous dimensions of the gluon and ghost fields as well as that of
the ghost-ghost-gluon vertex at four loops. All calculations have been
done in the general covariant gauge.

We apply our results to find the scheme and scale invariant gluon and
ghost propagators at Next-Next-Next-Leading Order (NNNLO) as well as
the four-loop anomalous dimension of the composite operator $A^2$.

\section{Notations and generalities}
The QCD Lagrangian with $n_f$  quark flavors in the covariant
gauge reads:
\begin{eqnarray}
{\cal L} & = & -\frac{1}{4} G^a_{\mu\nu} G^{a\mu\nu}
+\sum_{f=1}^{n_f} \bar \psi^f ( \mathrm{i}  \slash{D}  -m_f) \psi^f
\label{lagr:1}
\\
&-& \frac{1}{2\xi_L} (\partial^\mu A^a_\mu)^2
+ \partial^\mu \bar c ^a (\partial  c ^a - g f^{abc}  c ^b A^c_\mu)
{},
 \end{eqnarray} 
where
 \begin{eqnarray} 
G^a_{\mu\nu}  &=&  \partial_\mu A^a_\nu - \partial_\nu A^a_\mu +
g\,(A_\mu \times A_\nu)^a, \ \ \  
(A \times B)^a = f^{abc} A^b B^c,
\\
D_{\mu} &=& \partial_\mu - \mathrm{i} g A^a_\mu T^a, \ \ \ 
\slash{D} = \gamma^\mu D_\mu
{}.
 \end{eqnarray}  
%g f^{abc} A^a_\mu A^b_\nu, \quad
%[D_{\mu}]_{ij} = \delta_{ij}\partial_\mu - \mathrm{i} g A^a_\mu T^a_{ij}

The quark field $\psi^f_i$ has a mass $m_f$ and transforms as the
fundamental representation and the gluon fields $A^a_\mu$ as the
adjoint representation of the gauge group $ \mathrm{SU} (3)$. $T^a_{ij}$ and
$f^{abc}$ are the generators of the fundamental and adjoint
representation of the corresponding Lie algebra. The $ c ^a$ are the
ghost fields and $ \xi_L $ is the gauge parameter ($ \xi_L =0$ corresponds
to the Landau gauge).

By  adding to \re{lagr:1} all counterterms necessary to remove UV divergences 
from Green functions, one arrives at the bare 
QCD Lagrangian written in terms of the  renormalized fields\footnote{
For simplicity we set  the t${}'$ Hooft mass $\mu=1$ in  eq. (\ref{lagr:2})
below.
} :
\begin{eqnarray}
{\cal L}_0 & = &
- \frac{1}{4} Z_3\, ( \partial _{\mu}A_{\nu} -  \partial _{\nu}A_{\mu})^2
- \frac{1}{2}g\, Z_1^{3g} \, ( \partial _{\mu}A^a_{\nu} -  \partial _{\nu}A^a_{\mu})
\,  ( A_{\mu} \times A_{\nu})^a 
\nonumber
\\
&-&
 \frac{1}{4} g^2\, Z^{4g}_1\, ( A_\mu \times A_\nu)^2
- Z_\xi^{2g}\, \frac{1}{2  \xi_L }  ( \partial _\nu A_\mu)^2
+ Z^c_3\, \partial_\nu \bar c  \, (\partial_\nu c )
\label{lagr:2}
\\
&+& g\, Z_1^{ccg} \, \partial^\mu \bar c \, (A \times   c  )
+
Z_2\sum_{f=1}^{n_f} 
\bar \psi^f ( \mathrm{i}  \slash{ \partial } 
 + g Z^{\psi\psi g}_1 Z_2^{-1}\slash{A} - Z_m m_f )\, \psi^f
\nonumber
{}\,.
\end{eqnarray}
Here 
$Z_\xi^{2g}$ is expressed  through the RC of the gauge fixing parameter
$ \xi_L $ as follows
\[
\xi_{L,0} =Z_\xi  \xi_L , \ \ \ Z_\xi^{2g} = Z_3/Z_\xi
{}.
\]
$Z_3, Z_2,Z_3^{c}$ are the wave-function RCs
appearing in the relations between the renormalized and bare
gluon, quark and ghosts fields, viz.
 \begin{equation} 
A^{a \mu}_0 = \sqrt{Z_3}\ A^{a \mu},
\ \
\psi^f_0  = \sqrt{Z_2}\ \psi^f_0,
\ \
c^{a}_0   = \sqrt{Z_3^{c}}\ c^{a} 
{}.
 \end{equation} 
The full set of the vertex RCs
 \begin{equation} 
Z^V_1, \ \ \ V\in \{\mathrm{3g,\ 4g,\  c  c  g  , \ \psi  \psi g}\}
{}
 \end{equation} 
serve to renormalize 3-gluon, 4-gluon, ghost-ghost-gluon, 
quark-quark-gluon vertex functions respectively. 

The  Slavnov-Taylor identities allows one to express  all  four vertex
RCs in terms of an  independent one, $Z_g = \frac{ \displaystyle  g_0 \mu^{- \epsilon }}{ \displaystyle  g}$,
and the above listed wave function RCs.
The corresponding relations are:
 \begin{eqnarray} 
Z_\xi &=& Z_3, 
\label{WI:xi}
\\
Z_g &=& \sqrt{Z_1^{4g}} \,  (Z_3)^{-1}, \ \ 
\label{WI:4g}
\\
Z_g &=& Z_1^{3g} (Z_3)^{-3/2}, \ \ 
\label{WI:3g}
\\
Z_g &=& Z_1^{ccg} (Z_3)^{-1/2} (Z_3^c)^{-1}, \ \ 
\label{WI:ccg}
\\
Z_g &=& Z_1^{\psi\psi g} (Z_3)^{-1/2} (Z_2)^{-1}
\label{WI:qqg}
{}.
 \end{eqnarray} 

Within the $ \overline{\mbox{MS}} $ scheme each  RC does not  depend
on dimensional parameters (masses and momenta) and can be represented
as follows
 \begin{eqnarray} 
Z(h) = 1 + \sum_{n=1}^\infty \frac{z^{(n)}(h)}{ \epsilon ^n}
\label{}
{},
 \end{eqnarray} 
where $h = g^2/(16 \pi^2)$ and the parameter $ \epsilon $ is related to the running  
space time dimension $D$  via $D= 4 - 2 \epsilon $.
Given a RC $Z(h)$,  the  corresponding anomalous dimension is defined as
 \begin{equation} 
\gamma(h) = -\mu^2\frac{\mathrm{d} \log Z(h)}{\mathrm{d} \mu^2}
=h \frac{ \partial  z^{(1)}(h)}{ \partial  h}
= -\sum_{n=0}^\infty (\gamma)_n \, h^{(n+1)} 
\label{anom:dim:generic}
{}.
 \end{equation}   
Customarily, one also  defines $Z_h =Z_g^2$ and refers to the corresponding anomalous
dimension as the QCD $\beta$-function:
 \begin{equation} 
\beta(h) = 2 \gamma_g(h) =
2 h \, \frac{ \partial  z^{(1)}_g(h)}{ \partial  h}
=
 -\sum_{n=0}^\infty \beta_{n} h^{(n+1)}
\label{beta:function:generic}
{}.
 \end{equation} 
Eqs. (\ref{WI:4g}-\ref{WI:qqg}) imply that 
 \begin{eqnarray} 
\beta &=&    \, \gamma_1^{4g} - 2 \, \gamma_3,
\\
\beta &=& 2 \, \gamma_1^{3g} - 3 \, \gamma_3,
\\
\beta &=& 2  \, \gamma_1^{ccg} - 2 \  \gamma_3^{c} -  \, \gamma_3,
\label{ccg}
\\
\beta &=& 2  \, \gamma_1^{\psi\psi g} - 2 \  \gamma_2 -  \, \gamma_3
\label{qqg}
{}.
 \end{eqnarray}

As is well-known there is  a one-to-one correspondence between an anomalous dimension
and the corresponding RC. For instance, 
$Z_h$ obeys an   equation
 \begin{equation} 
\left(- \epsilon  \,   + \beta(h)\right)\, h \, \frac{ \partial  \log Z_h}{ \partial  h} = -\beta(h)
\label{Z_h from beta}
 \end{equation} 
which leads to 
\[
\log Z_h=   \displaystyle  \int \frac{\mathrm{d} h}{h} \, \frac{\beta}{ \epsilon  \,  - \beta} 
{}.
\]
In general case  $Z$  depends on $\mu$  
through both  $h$ and $ \xi_L $ and  an analog of  eq.~\re{Z_h from beta} assumes the form:
 \begin{equation} 
\left(- \epsilon   + \beta(h)\right)\, h \, \frac{ \partial  \log Z }{ \partial  h} + \gamma_3(h) 
\,  \xi_L  \, \frac{ \partial  \log Z}{ \partial   \xi_L } = -\gamma(h)
\label{Z from beta}
{}.
 \end{equation} 
Eq. \re{Z from beta} can be easily utilized   to reconstruct $Z$ from
$\gamma$ and $\beta$. As anomalous dimensions are more compact  than  corresponding RCs
in what follows we will  write explicitly only the former.

%An inspection of \ref{} shows 

\section{Calculations and Results}

Relations (\ref{WI:4g}-\ref{WI:qqg}) demonstrate that a minimal set of
the RCs  necessary to reconstruct all
coefficients of the bare QCD Lagrangian \re{lagr:2} consists of
$Z_m$, all three wave-function RCs  $Z_3, Z_2,
Z_3^{c}$, and the coupling constant RC $Z_g$ or, instead, at least one
from the collection  $Z^V_1, \ \ \ V\in \{\mathrm{3g,\ 4g,\ c c g, \ 
\psi \psi g}\}$. Taking into account that $Z_m, Z_2$ and $Z_g$ are known with
four loop accuracy from the works 
\cite{vanRitbergen:1997va,Chetyrkin:1997dh,Vermaseren:1997fq,Chetyrkin:1999pq},
one is left with just two
specific RC to compute, say, $Z_3$ and $Z_3^{c}$.

At present there are basically two different ways to perform RG
calculations at the four-loop level. Both approaches make use of the
method of Infrared Rearrangement (IRR) Ref.~\cite{Vladimirov:1980zm} in
order to set zero (possibly after a proper Taylor expansion) masses
and external momenta. Both eventually employ the traditional integration
by parts method to compute the resulting Feynman
integrals\footnote{Though the technical implementations could be
quite different, cf. works 
\cite{Laporta:1996mq,Chetyrkin:1998fm,Mastrolia:2000va,Laporta:2001dd,Schroder:2002re}.}.

The first one, pioneered in the yearly works of Dubna group 
\cite{Tarasov:1977ef,Kazakov:1979ik,Tarasov:1980au}, 
amounts to adding an artificial mass or an  external momentum to a
properly chosen  propagator  of a given Feynman diagram before the expansion
in masses and true external momenta is made. The artificial external
momentum has to be introduced in such a way that all spurious infrared
divergences are removed and the obtained Feynman integral is
calculable. In practical multiloop calculations the condition of
absence of the infrared divergences leads to unnecessary complications
and, in some cases, even prevents from reduction to the simplest
integrals. The problem was solved  with elaborating a
special technique of subtraction of IR divergences --- the
$R^*$-operation \cite{ChS:R*,gssq,gvvq}. This technique succeeds in
expressing the UV counterterm of every (L+1)-loop Feynman integral in
terms of divergent and finite parts of some L-loop massless
propagators.

In the second approach the infrared rearrangement is performed by
introducing a {\em single} auxiliary mass to {\em all} propagators in
each Feynman diagram at hand 
\cite{Misiak:1995zw,Chetyrkin:1997vx,vanRitbergen:1997va}. 
No IR divergences can ever appear due to absence of any massless
propagators. Next, after a proper expansion in all the particle masses
(except the auxiliary one) and external momenta is performed. The
resulting integrals are completely massive tadpoles, i.e. Feynman
integrals without external momenta and with only a single mass
inserted in all the propagators.

In our calculation of $Z_3$ and $Z_3^{c}$ we have used the first, 
``massless'' approach. It proved also  to be more convenient to
compute the RC $Z_1^{ccg}$ instead of $Z_3$ and then to find  $Z_3$ from
eq. \re{WI:ccg}.

The four-loop diagrams contributing to the ghost propagator and to the
ghost-ghost-gluon vertex to order $\alpha_s^4$ (altogether about
35000) have been generated with the program QGRAF
\cite{qgraf}, then  globally rearranged to a product of
some three-loop p-integrals with a trivial (essentially one-loop)
massive Feynman integral and, finally, computed with the program
MINCER \cite{Vermaseren:91,mincer:91}. The total amount of CPU time
needed to compute RC $Z_3^{c}$ and $Z_1^{ccg}$ was about a month of
work of a standard PC with an Athlon XP 2000+ processor.  For testing
purposes we have also computed the RC $Z_1^{qqg}$, which has required
an almost double amount of calculational time\footnote{These figures
should be understood as {\em effective} ones; that is in reality we
have used a sort of trivial parallelization by distributing diagrams
between a few PC's.}.

Our results for the  anomalous dimensions $\gamma_1^{ccg}$   and $\gamma^c_3$  read
%\input{RG41}

%%%%%%%%%%%%%%%%%%%%%%%%%%%%%%%%%%%%%%

\begin{eqnarray}
(\gamma_1^{ccg})_0 =  
\frac{3}{2}  \, \xi_L 
{},
\label{g_ccg_0}
\end{eqnarray}
%zero == 0
\begin{eqnarray}
(\gamma_1^{ccg})_1 =  
{} 
\frac{45}{8}  \, \xi_L 
+\frac{9}{8}  \, \xi_L ^2
{},
\label{g_ccg_1}
\end{eqnarray}
%zero == 0
\begin{eqnarray}
(\gamma_1^{ccg})_2 =  
&{}& 
 \frac{5427}{64}  \, \xi_L 
+\frac{1053}{64}  \, \xi_L ^2
+\frac{135}{32}  \, \xi_L ^3
-\frac{135}{16}  \, \xi_L \,n_f
{},
\label{g_ccg_2}
\end{eqnarray}
%zero == 0
\begin{eqnarray}
\lefteqn{(\gamma_1^{ccg})_3 =  } 
\nonumber\\
&{}& 
\left.
\frac{635749}{384}  \, \xi_L 
+\frac{21519}{32}  \,\zeta_{3} \, \xi_L 
+\frac{729}{64}  \,\zeta_{4} \, \xi_L 
-\frac{91125}{128}  \,\zeta_{5} \, \xi_L 
+\frac{29547}{128}  \, \xi_L ^2
  \right. \nonumber \\ &{}& \left.  
\phantom{+ }
+\frac{6993}{64}  \,\zeta_{3} \, \xi_L ^2
+\frac{243}{16}  \,\zeta_{4} \, \xi_L ^2
-\frac{8505}{128}  \,\zeta_{5} \, \xi_L ^2
+\frac{7371}{128}  \, \xi_L ^3
+\frac{729}{16}  \,\zeta_{3} \, \xi_L ^3
  \right. \nonumber \\ &{}& \left.  
\phantom{+ }
+\frac{243}{64}  \,\zeta_{4} \, \xi_L ^3
-\frac{6075}{128}  \,\zeta_{5} \, \xi_L ^3
+\frac{1539}{128}  \, \xi_L ^4
-\frac{459}{64}  \,\zeta_{3} \, \xi_L ^4
+\frac{945}{128}  \,\zeta_{5} \, \xi_L ^4
%zero == 0
\right.
\nonumber\\
&{+}& \,n_f 
\left[
-\frac{54623}{288}  \, \xi_L 
-\frac{663}{8}  \,\zeta_{3} \, \xi_L 
-\frac{99}{4}  \,\zeta_{4} \, \xi_L 
-\frac{453}{32}  \, \xi_L ^2
-\frac{45}{8}  \,\zeta_{3} \, \xi_L ^2
%zero == 0
\right]
\nonumber\\
&{+}& \, n_f^2
\left[
-\frac{251}{54}  \, \xi_L 
+6  \,\zeta_{3} \, \xi_L 
%zero == 0
\right]
{},
\label{g_ccg_3}
\end{eqnarray}
%zero == 0
%%%%%%%%%%%%%%%%%%%%%%%%%%%%%%%%%%%%%%
\begin{eqnarray}
(\gamma_3^c)_0 =   
-\frac{9}{4} 
+\frac{3}{4}  \, \xi_L 
{},
\label{g_cc_0}
\end{eqnarray}
%zero == 0
\begin{eqnarray}
(\gamma_3^c)_1 =   
-\frac{285}{16} 
-\frac{9}{16}  \, \xi_L 
%zero == 0
{+}
 \frac{5}{4} \,n_f 
{},
\label{g_cc_1}
\end{eqnarray}
%zero == 0
\begin{eqnarray}
(\gamma_3^c)_2 &=&   
%\nonumber\\
\left.
-\frac{15817}{64} 
-\frac{243}{32}  \,\zeta_{3}
+\frac{459}{32}  \, \xi_L 
-\frac{81}{8}  \,\zeta_{3} \, \xi_L 
+\frac{81}{32}  \, \xi_L ^2
  \right. \nonumber \\ &{}& \left.  
\phantom{+ }
-\frac{81}{32}  \,\zeta_{3} \, \xi_L ^2
+\frac{81}{64}  \, \xi_L ^3
%zero == 0
\right.
\nonumber\\
&{+}& \,n_f 
\left[
\frac{637}{24} 
+\frac{33}{2}  \,\zeta_{3}
-\frac{63}{16}  \, \xi_L 
%zero == 0
\right]
{+} 
 \frac{35}{36} \, n_f^2
{},
\label{g_cc_2}
\end{eqnarray}
%zero == 0
\begin{eqnarray}
\lefteqn{(\gamma_3^c)_3 =  } 
\nonumber\\
&{}& 
\left.
-\frac{2857419}{512} 
-\frac{1924407}{512}  \,\zeta_{3}
+\frac{8019}{64}  \,\zeta_{4}
+\frac{40905}{8}  \,\zeta_{5}
+\frac{368231}{1536}  \, \xi_L 
  \right. \nonumber \\ &{}& \left.  
\phantom{+ }
-\frac{75573}{256}  \,\zeta_{3} \, \xi_L 
+\frac{17901}{128}  \,\zeta_{4} \, \xi_L 
-\frac{12015}{256}  \,\zeta_{5} \, \xi_L 
+\frac{17613}{512}  \, \xi_L ^2
-\frac{4131}{128}  \,\zeta_{3} \, \xi_L ^2
  \right. \nonumber \\ &{}& \left.  
\phantom{+ }
+\frac{81}{2}  \,\zeta_{4} \, \xi_L ^2
-\frac{21465}{256}  \,\zeta_{5} \, \xi_L ^2
+\frac{4185}{512}  \, \xi_L ^3
+\frac{9045}{256}  \,\zeta_{3} \, \xi_L ^3
+\frac{729}{128}  \,\zeta_{4} \, \xi_L ^3
  \right. \nonumber \\ &{}& \left.  
\phantom{+ }
-\frac{17145}{256}  \,\zeta_{5} \, \xi_L ^3
+\frac{81}{32}  \, \xi_L ^4
-\frac{3213}{512}  \,\zeta_{3} \, \xi_L ^4
+\frac{945}{256}  \,\zeta_{5} \, \xi_L ^4
%zero == 0
\right]
\nonumber\\
&{+}& \,n_f 
\left[
\frac{1239661}{1152} 
+\frac{48857}{48}  \,\zeta_{3}
-\frac{8955}{32}  \,\zeta_{4}
-\frac{3355}{4}  \,\zeta_{5}
-\frac{11107}{288}  \, \xi_L 
  \right. \nonumber \\ &{}& \left.  
\phantom{+ \,n_f }
-\frac{39}{8}  \,\zeta_{3} \, \xi_L 
-\frac{153}{8}  \,\zeta_{4} \, \xi_L 
-\frac{651}{128}  \, \xi_L ^2
-\frac{27}{8}  \,\zeta_{3} \, \xi_L ^2
-\frac{27}{32}  \,\zeta_{4} \, \xi_L ^2
%zero == 0
\right]
\nonumber\\
&{+}& \, n_f^2
\left[
-\frac{586}{27} 
-\frac{55}{2}  \,\zeta_{3}
+\frac{33}{2}  \,\zeta_{4}
-\frac{779}{432}  \, \xi_L 
+3  \,\zeta_{3} \, \xi_L 
%zero == 0
\right]
\nonumber\\
&{+}& \, n_f^3
\left[
\frac{83}{108} 
-\frac{4}{3}  \,\zeta_{3}
%zero == 0
\right]
{}.
\label{g_cc_3}
\end{eqnarray}
%zero == 0

Finally, a use of eq.~ \re{ccg} immediately leads us  to

\begin{eqnarray}
(\gamma_3)_0 =   
-\frac{13}{2} 
+\frac{3}{2}  \, \xi_L 
%zero == 0
{+} \,n_f 
 \frac{2}{3}
{},
\label{g3_0}
\end{eqnarray}
%zero == 0
\begin{eqnarray}
(\gamma_3)_1 =   
-\frac{531}{8} 
+\frac{99}{8}  \, \xi_L 
+\frac{9}{4}  \, \xi_L ^2
%zero == 0
{+} 
 \frac{61}{6} \,n_f 
{},
\label{g3_1}
\end{eqnarray}
%zero == 0
\begin{eqnarray}
\lefteqn{(\gamma_3)_2 =  } 
\nonumber\\
&{}& 
\left.
-\frac{29895}{32} 
+\frac{243}{16}  \,\zeta_{3}
+\frac{4509}{32}  \, \xi_L 
+\frac{81}{4}  \,\zeta_{3} \, \xi_L 
+\frac{891}{32}  \, \xi_L ^2
  \right. \nonumber \\ &{}& \left.  
\phantom{+ }
+\frac{81}{16}  \,\zeta_{3} \, \xi_L ^2
+\frac{189}{32}  \, \xi_L ^3
%zero == 0
\right.
\nonumber\\
&{+}& \,n_f 
\left[
\frac{8155}{36} 
-33  \,\zeta_{3}
-9  \, \xi_L 
%zero == 0
\right]
-
\frac{215}{27}\, n_f^2
{},
\label{g3_2}
\end{eqnarray}
%zero == 0
\begin{eqnarray}
\lefteqn{(\gamma_3)_3 =   }
\nonumber\\
&{}& 
\left.
-\frac{10596127}{768} 
+\frac{1012023}{256}  \,\zeta_{3}
-\frac{8019}{32}  \,\zeta_{4}
-\frac{40905}{4}  \,\zeta_{5}
+\frac{2174765}{768}  \, \xi_L 
  \right. \nonumber \\ &{}& \left.  
\phantom{+ }
+\frac{247725}{128}  \,\zeta_{3} \, \xi_L 
-\frac{16443}{64}  \,\zeta_{4} \, \xi_L 
-\frac{170235}{128}  \,\zeta_{5} \, \xi_L 
+\frac{100575}{256}  \, \xi_L ^2
+\frac{18117}{64}  \,\zeta_{3} \, \xi_L ^2
  \right. \nonumber \\ &{}& \left.  
\phantom{+ }
-\frac{405}{8}  \,\zeta_{4} \, \xi_L ^2
+\frac{4455}{128}  \,\zeta_{5} \, \xi_L ^2
+\frac{25299}{256}  \, \xi_L ^3
+\frac{2619}{128}  \,\zeta_{3} \, \xi_L ^3
-\frac{243}{64}  \,\zeta_{4} \, \xi_L ^3
  \right. \nonumber \\ &{}& \left.  
\phantom{+ }
+\frac{4995}{128}  \,\zeta_{5} \, \xi_L ^3
+\frac{1215}{64}  \, \xi_L ^4
-\frac{459}{256}  \,\zeta_{3} \, \xi_L ^4
+\frac{945}{128}  \,\zeta_{5} \, \xi_L ^4
%zero == 0
\right.
\nonumber\\
&{+}& \,n_f 
\left[
\frac{23350603}{5184} 
-\frac{387649}{216}  \,\zeta_{3}
+\frac{8955}{16}  \,\zeta_{4}
+\frac{3355}{2}  \,\zeta_{5}
-\frac{10879}{36}  \, \xi_L 
  \right. \nonumber \\ &{}& \left.  
\phantom{+ \,n_f }
-156  \,\zeta_{3} \, \xi_L 
-\frac{45}{4}  \,\zeta_{4} \, \xi_L 
-\frac{1161}{64}  \, \xi_L ^2
-\frac{9}{2}  \,\zeta_{3} \, \xi_L ^2
+\frac{27}{16}  \,\zeta_{4} \, \xi_L ^2
%zero == 0
\right]
\nonumber\\
&{+}& \, n_f^2
\left[
-\frac{43033}{162} 
-\frac{2017}{81}  \,\zeta_{3}
-33  \,\zeta_{4}
-\frac{1229}{216}  \, \xi_L 
+6  \,\zeta_{3} \, \xi_L 
%zero == 0
\right]
\nonumber\\
&{+}& \, n_f^3
\left[
-\frac{4427}{1458} 
+\frac{8}{3}  \,\zeta_{3}
%zero == 0
\right]
{}.
\label{g3_3}
\end{eqnarray}
%zero == 0
For an important particular case of the Landau gauge $ \xi_L  = 0$ we get:
\begin{eqnarray}
(\gamma_3)_0 =   
-\frac{13}{2} 
%zero == 0
{+} \,n_f 
 \frac{2}{3}
{},
\ \ \ 
(\gamma_3)_1 =   
-\frac{531}{8} 
%zero == 0
{+} \,n_f 
 \frac{61}{6}
{},
\label{g3_1:Landau}
\end{eqnarray}
%zero == 0
 \begin{equation} 
(\gamma_3)_2 =  
-\frac{29895}{32} 
+\frac{243}{16}  \,\zeta_{3}
%zero == 0
{+} \,n_f 
\left[
\frac{8155}{36} 
-33  \,\zeta_{3}
%zero == 0
\right]
-\, n_f^2
\frac{215}{27}
{},
\label{g3_2:Lanadau}
 \end{equation} 
\begin{eqnarray}
(\gamma_3)_3 =  
&-&\frac{10596127}{768} 
+\frac{1012023}{256}  \,\zeta_{3}
-\frac{8019}{32}  \,\zeta_{4}
-\frac{40905}{4}  \,\zeta_{5}
%zero == 0
\nonumber\\
&{+}& \,n_f 
\left[
\frac{23350603}{5184} 
-\frac{387649}{216}  \,\zeta_{3}
+\frac{8955}{16}  \,\zeta_{4}
+\frac{3355}{2}  \,\zeta_{5}
%zero == 0
\right]
\nonumber\\
&{+}& \, n_f^2
\left[
-\frac{43033}{162} 
-\frac{2017}{81}  \,\zeta_{3}
-33  \,\zeta_{4}
%zero == 0
\right]
\nonumber\\
&{+}& \, n_f^3
\left[
-\frac{4427}{1458} 
+\frac{8}{3}  \,\zeta_{3}
%zero == 0
\right]
{}.
\label{g3_3:Lanadau}
\end{eqnarray}
%zero == 0
or, numerically,
 \begin{eqnarray} 
\gamma_3 &=& h (6.5 - 0.666667 n_f) + h^2  (66.375 - 10.1667 n_f) 
\nonumber
\\
&{+}& h^3  (915.963 - 186.86 n_f + 7.96296 n_f^2 ) 
\nonumber
\\
&{+}& h^4  (19920.2 -  4692.27 n_f + 331.285 n_f^2  - 0.169134 n_f^3 )
\label{gamma_3:numerics}
{}.
 \end{eqnarray}

\section{Applications} 
In this section we consider some applications of our results. 
The case of the massless QCD with the Landau gauge fixing is
understood in both subsections.

\subsection{Scheme-invariant Gluon and Ghost Propagators in NNNLO}

In general case a (multiplicatively renormalizable) Green function $G$
depends on both a renormalization prescription ({\em scheme}) and the
choice of the normalization scale $\mu$. In many cases it is more
convenient to deal with the scheme and scale invariant version of $G$
which we will denote as $\hat{G}$. Given the RG equation for $G$ 
 \begin{equation} 
\mu^2 \frac{\mathrm{d}}{\mathrm{d} \mu^2}
G(h,\mu) \equiv
\left(
\mu^2\frac{\partial}{\partial\mu^2}
 +
\beta(h)
h
\frac{\partial}{\partial h}
\right)
      G          = \gamma(h)\, G(h,\mu) 
\label{rg:Pi2}
{},
 \end{equation} 
then a formal solution for $\hat{G}$ reads
 \begin{equation}  
\hat{G}  = 
 G(h,\mu)/  f(h)
,
\ \ 
f(h)  = 
\mathrm{exp} \left\{
\int^h \,\frac{ \mathrm{d} x}{x}\ \frac{\gamma(x)}{\beta(x)}
\right\}
\label{formal_sol}
{},
 \end{equation} 
 \begin{eqnarray}  
\nonumber
f(h) &=& (h)^{\bar{ \gamma _0}}  \left\{ 1 + (\bar{ \gamma _1} - \bar{\beta_1}\bar{ \gamma _0})h
\right.
\\ \nonumber
&+&
\frac{1}{2}
\left[
(\bar{ \gamma _1} - \bar{\beta_1}\bar{ \gamma _0})^2
+
\bar{ \gamma _2} + \bar{\beta_1}^2\bar{ \gamma _0}
- \bar{\beta_1}\bar{ \gamma _1} -\bar{\beta_2}\bar{ \gamma _0}
\right] h^2
\\ \label{c(h)}
&+&
\left[
\frac{1}{6}(\bar{ \gamma _1} - \bar{\beta_1}\bar{ \gamma _0})^3
+
\frac{1}{2}(\bar{ \gamma _1} - \bar{\beta_1}\bar{ \gamma _0})
(
\bar{ \gamma _2} + \bar{\beta_1}^2\bar{ \gamma _0}
- \bar{\beta_1}\bar{ \gamma _1} -\bar{\beta_2}\bar{ \gamma _0}
)
\right.
\\ \nonumber
&&
+\frac{1}{3}\left(
\left.\left.
\bar{ \gamma _3}
-\bar{\beta_1}^3\bar{ \gamma _0} + 2\bar{\beta_1} \bar{\beta_2}\bar{ \gamma _0}
-\bar{\beta_3}\bar{ \gamma _0} + \bar{\beta_1^2}\bar{ \gamma _1}
- \bar{\beta_2}\bar{ \gamma _1} - \bar{\beta_1}\bar{ \gamma _2}
\right)
\right] h^3 + {\cal O}(h^4)
\right\}
{}.
 \end{eqnarray} 
Here $\bar{ \gamma _i} =  \gamma _i/\beta_0$, $\bar{\beta_i} = \beta_i/\beta_0$,
(i=1,2,3), the coefficients $\gamma_i, \  \beta_i$ are defined in 
eqs.~(\ref{anom:dim:generic},\ref{beta:function:generic}).

Now, after  writing  the gluon 
and ghost propagators in the form
 \begin{equation} 
D^{ab}_{\mu\nu}(q) = \frac{\delta^{ab}}{-q^2}
\left[
 -g_{\mu\nu} + \frac{q_{\mu} q_{\nu}}{q^2}
\right] D(-q^2)
{},
\ \ 
\Delta^{ab}(q) = \frac{\delta^{ab}}{-q^2}
\ \ 
 \Delta(-q^2)
\label{gluon_ghost_prop}
{}
 \end{equation} 
and a use of explicit expressions for the very propagators from \cite{Chetyrkin:2000dq}
we arrive at the following NNNLO  predictions for the asymptotic behavior of the
scheme and scale invariant functions $\hat{D}$ and $\hat{\Delta}$ at large Euclidean $Q^2 =-q^2$
in the $ \overline{\mbox{MS}} $ scheme. First, for the case of pure gluodynamics ($n_f =0$)

\begin{eqnarray}
&{}& h^{\frac{13}{22}}\,\, \hat{D}^{-1}(-q^2)|_{n_f=0} = 
1{-}
\frac{25085}{2904}\, h 
{+}\, h^2 
\left[
-\frac{412485993}{1874048} 
+\frac{9747}{352}  \,\zeta_{3}
%zero == 0
\right]
\nonumber\\
&{+}&\, h^3 
\left[
-\frac{141629801206331}{16326706176}
+ \frac{80968605}{42592} \,\zeta_{3}
+\frac{477315}{704}  \,\zeta_{5}
%zero == 0
\right]
{}
\label{SchemeInvGluonProp}
\nonumber\\
&{=}& 
1
{-}\,8.63809 h 
{-} 186.819\, h^2 
{-} 5686.55\, h^3 
{},
\label{SchemeInvGluonPropN}
\end{eqnarray}
%zero == 0. + 0.*a + 0.*a^2 + 0.*a^3
\begin{eqnarray}
\lefteqn{
h^{\frac{9}{44}}\,\, 
\hat{\Delta}^{-1}(-q^2)|_{n_f=0} =  
%\nonumber\\
%&{+}& 
1
{-}
\frac{5271}{1936}\, h 
{+}\, h^2 
\left[
-\frac{615512003}{7496192} 
+\frac{5697}{704}  \,\zeta_{3}
%zero == 0
\right]
}
\nonumber\\
&{+}&\, h^3 
\left[
-\frac{430343889400537}{130613649408}
+\frac{674654895}{1362944}   \,\zeta_{3}
+\frac{73845}{352}  \,\zeta_{5}
%zero == 0
\right]
\nonumber\\
&{=}& 
1
{-} 2.72262\, h 
{-}72.3825 \, h^2 
{-} 2482.24\, h^3 
{}.
\label{SchemeInvGhostProp}
\end{eqnarray}
%zero == 0

In order to illustrate the $n_f$ dependence we give below the results for $n_f=3$ and $n_f=6$ (to save
space only in the numerical form) 
 \begin{equation} 
h^{\frac{1}{2}}\,\, 
\hat{D}^{-1}(-q^2)|_{n_f=3} =  1 - 5.18056 \, h - 85.0853 \, h^2  - 2178.1 \, h^3
\label{gluon:nf:3}
{},
 \end{equation} 
 \begin{equation} 
h^{\frac{1}{4}}\,\, 
\hat{\Delta}^{-1}(-q^2)|_{n_f=3} =   1 - 2.78472 \, h - 52.591 \, h^2  -1359.11 \, h^3
\label{ghost:nf:3}
{},
 \end{equation} 

 \begin{equation} 
h^{\frac{5}{14}}\,\, 
\hat{D}^{-1}(-q^2)|_{n_f=6} =  1  - 0.857993 \, h + 16.6153 \, h^2  +  220.455 \, h^3
\label{gluon:nf:6}
{},
 \end{equation} 
 \begin{equation} 
h^{\frac{9}{28}}\,\, 
\hat{\Delta^{-1}}(-q^2)|_{n_f=6} =   1 - 3.27934 \, h - 31.5908 \, h^2  - 372.071\, h^3
\label{ghost:nf:6}g
{}.
 \end{equation} 

Note that  in eqs.~(\ref{SchemeInvGluonProp}-\ref{ghost:nf:6}) 
the coupling constant $h$ should be understood as $h(\mu^2 = -q^2)$.

\subsection{Four-Loop Anomalous Dimension of the Composite Operator $A^2$}

Recently there has been a lot of activity in studying the possibility
of a condensate in Yang-Mills-theory of mass dimension two (see,
e.g. recent works \cite{Dudal:2003by,Esole:2004jd} and references therein). 
The relevance of the operator $A^2 \equiv A^a_{\mu} A^a_{\mu}$ in the Landau gauge in
that context has been widely discussed. In this connection
a thorough investigation of the renormalization properties of the
composite operator $A^2 \equiv A^a_{\mu} A^a_{\mu}$ has been carried
out in \cite{Gracey:2002yt,Dudal:2003np}.  
In particular, the author of \cite{Gracey:2002yt}  has 
discovered by  explicit three loop computation
a remarkable relation\footnote{We have adjusted the coefficients in
(\ref{exact1}) to our notations.}
 \begin{equation} 
-2 \gamma_{A^2}|_{\xi_L =0} = \beta - \gamma_3
\label{exact1}
 \end{equation} 
expressing the anomalous dimension $\gamma_{A^2}$ of the operator $A^2$
in terms of the $\beta$-function and the gluon field anomalous
dimension $\gamma_3$. An all  orders proof of \re{exact1} have been later constructed in 
\cite{Dudal:2002pq} using algebraic renormalization methods.
As both ingredients
of \re{exact1} are known now at four loops one arrives at

\begin{eqnarray}
\lefteqn{\gamma_{A^2}|_{\xi_L =0} = 
 h 
\left[
\frac{35}{4} 
-\frac{2}{3}  \,n_f 
\right]
{+}\, h^2 
\left[
\frac{1347}{16} 
-\frac{137}{12}  \,n_f 
%zero == 0
\right]}
\nonumber\\
&{+}&\, h^3 
\left[
\frac{75607}{64} 
-\frac{243}{32}  \,\zeta_{3}
+  \,n_f \left(
-\frac{18221}{72} 
+\frac{33}{2}    \,\zeta_{3}
     \right)
+\frac{755}{108}  \, n_f^2
%zero == 0
\right]
\nonumber\\
&{+}&\, h^4
\left[
\frac{29764511}{1536} 
-\frac{99639}{512}  \,\zeta_{3}
+\frac{8019}{64}  \,\zeta_{4}
+\frac{40905}{8}  \,\zeta_{5}
  \right. \nonumber \\ &{}& \left.  
\phantom{+\, h^4}
+  \,n_f \left(
-\frac{57858155}{10368}  
+\frac{335585}{432}    \,\zeta_{3}
-\frac{3355}{4}   \,\zeta_{5}
-\frac{8955}{32}  \,\zeta_{4}
 \right)
  \right. \nonumber \\ &{}& \left.  
\phantom{+\, h^4}
+  \,n_f^2 \left(
\frac{46549}{162}  
+\frac{8489}{162}   \,\zeta_{3}
+\frac{33}{2}   \,\zeta_{4}
 \right)
  \right. \nonumber \\ &{}& \left.  
\phantom{+\, h^4}
+  \,n_f^3 \left(
\frac{6613}{2916}  
-\frac{4}{3}  \, \zeta_{3}
 \right)
%zero == 0
\right]
{}.
\label{anom[A^2]}
\end{eqnarray}
%zero == 0

\section{Discussion and Conclusion} 

Calculation of the two missing RCs $Z_3$ and $Z^c_2$   completes  
renormalization of the QCD Lagrangian at four loops. An important issue relevant
for any calculation of such complexity is  the correctness of the obtained anomalous dimensions.
The following comments are in order.

\begin{itemize}

\item The FORM program MINCER used by us to compute 
three loop massless propagators was developed more than a decade ago
and has been since heavily cross-checked in a number  of various
multiloop calculations.

\item  At three loop level we have  full agreement with the results of 
\cite{Larin:1993tp,Chetyrkin:1999pq}. 
This checks our way to  use $R^*$-operation because  the  three loop results of 
\cite{Larin:1993tp,Chetyrkin:1999pq} have been obtained with direct application of MINCER to 
three-loop propagators. In the present work the {\em three loop}  contributions come
exclusively  from  {\em two loop} propagators.

\item  The  leading $n_f$ behaviour of the quark, gluon and ghost anomalous dimensions 
as well as the anomalous dimensions of the quark-quark-gluon and
ghost-ghost-gluon vertices  was  investigated in all orders of perturbation theory 
in the work \cite{Gracey:1993ua}. At the four loop level the predictions  of 
\cite{Gracey:1993ua} are in full agreement to the corresponding leading $n_f$ pieces of 
our results\footnote{We thank John Gracey for this comment.}.

\item We have also computed the ${\cal{O}}(\alpha_s^4)$ 
anomalous dimension of the quark-quark-gluon vertex 
and found that it satisfies  eq. \re{qqg} as it should.

\item 
As is known from \cite{Taylor:1971,Blasi:1991xz} the ghost-ghost-gluon
vertex is unrenormalized in the Landau gauge, that is
 \begin{equation} 
\gamma_{1}^{ccg}|_{\xi_L =0} =0  
\label{gamma_gcc:landau}
{}.
 \end{equation} 	
Eq.  \re{gamma_gcc:landau} is in obvious agreement to  eqs.~(\ref{g_ccg_0} - \ref{g_ccg_3}).

\item We have {\em not} performed a {\em direct} calculation 
of the gluon wave function RC $Z_3$ but rather extract the result from
the $\beta$-function of
\cite{vanRitbergen:1997va} and the  anomalous dimensions
$\gamma_3^c$ and $\gamma_1^{ccg}$.  Thus, an independent reevaluation
of either $\gamma_3$ or/and the $\beta$-function at four loops is
highly desirable.

\end{itemize}

An interesting feature of the scheme and scale invariant functions
$\hat{D}$ and $\hat{\Delta}$ is the full absence of the irrational
constant $\zeta_4$ as illustrated in
eqs.~(\ref{SchemeInvGluonProp},\ref{SchemeInvGhostProp}).  It stems
from a noni-trivial mutual cancellation of the terms proportional to
$\zeta_4$ which enter into the ingredients---the Green function and
its anomalous dimension---of the definition of $\hat{G}$ (see
eq.~\re{formal_sol}). We have checked that this is true also for the
$SU(N)$ color group and a generic value of $n_f$.

Finally, a comment about the gauge group dependence of our results. Preferring
compact and readable formulas to huge and general ones, the author has
deliberately formatted all results in this paper for the most
practically important case of the $SU(3)$ color group. In reality all
calculations have been carried out for a little bit more general case
of the SU(N) color group.  Full expressions of  RCs (and the 
corresponding anomalous dimensions) describing the renormalization of
the Lagrangian \re{lagr:2} in the general covariant gauge and for the SU(N)
color group  are  available  (in a computer-readable  form)   
in http://www-ttp.physik.uni-karlsruhe.de/Progdata/ttp04/ttp04-08/.

{\it Acknowledgments:} I would like to thank Michael~Czakon and York~Schr\"oder for
helpful discussions which have motivated me  to finish  the  old project.
I also want to thank Mboyo Esole for information about  the work 
\cite{Esole:2004jd}.

This work was supported by the Deutsche Forschungsgemeinschaft 
in the Sonderforschungsbereich/Transregio SFB/TR-9 ``Computational
 Particle Physics'', by Volkswagen Foundation and
by the European Union under contract HPRN-CT-2000-00149.

{\it Note 1}

Just before the  submission of the manuscript for publication we have been informed  
that our main assumption --- the validity of the result for four-loop
QCD beta-function first obtained in \cite{vanRitbergen:1997va} ---
is confirmed by a completely independent calculation \cite{Czakon200}.

\end{document}